\documentclass[aps,prl,reprint,showpacs,superscriptaddress]{revtex4-1}

\usepackage{amsmath,amssymb,bm}
\usepackage{hyperref}
\def\pa{\partial_\alpha}
\def\pb{\partial_\beta}
\def\al{\alpha}
\def\be{\beta}
\def\dps{{d}^{p+1}\!\sigma\:}
\def\dX{\widetilde{\partial\! X}{}}
\DeclareMathOperator{\tr}{\text{tr}}

\DeclareMathOperator{\sgn}{\text{sgn}}

\begin{document}

%\preprint{}

\title{Nambu-Sigma model and effective membrane actions}

\author{Branislav Jur\v co}
\affiliation{Mathematical Institute, Charles University, Prague 186 75, Czech Republic}

\author{Peter Schupp}
%\email[]{p.schupp@jacobs-university.de}
\affiliation{Jacobs University Bremen, 28759 Bremen, Germany}
\altaffiliation[Current address: ]{Maxwell Institute for Mathematical Sciences,
Department of Mathematics, Heriot-Watt University, Edinburgh, EH14 4AS, Scotland}

\date{\today}

\begin{abstract}
We propose an effective action for a $p'$-brane with open $p$-branes ending
on it. The action has dual descriptions similar to the commutative and non-commutative ones of
the DBI action for \mbox{D-}branes and open strings.
The Poisson structure governing the non-commutativity of the D-brane
is replaced by a Nambu structure and the open-closed string relations are
generalized to the case of $p'$-branes utilizing a novel Nambu sigma model
description of $p$-branes.
In the case of an M5-brane our action interpolates between M5-actions
already proposed in the literature and matrix model like actions
involving Nambu structures.
\end{abstract}


\maketitle

\emph{Introduction.}---%
The construction of an all-order effective action for the M5 brane with multiple open M2 branes
has been an open problem for well over a decade. For a single D-brane in open superstring theory, the Dirac-Born-Infeld action provides such an effective description valid to all orders \cite{Fradkin:1985ys,Leigh:1989jq,Tseytlin:1999dj}. In fact there are equivalent commutative and non-commutative descriptions \cite{Seiberg:1999vs} and this equivalence is quite restrictive. It has been used 
to study the non-abelian DBI action \cite{Cornalba:2000ua,Terashima:2000ej} and more 
recently in the context of the M2/M5 brane system \cite{Chen:2010br}. We pick up on this idea, introduce a Nambu sigma model, determine open-closed membrane relations and a Nambu-Poisson map that relates ordinary higher gauge theory to a new type of Nambu gauge theory. We find that these considerations fix the bosonic part of the desired effective action essentially uniquely. Imposing $\kappa$-symmetry it should be possible to determine the full supersymmetric action, but we shall focus on the bosonic part in this letter.
Our action interpolates between early proposals from the 1990s and matrix-model like actions involving Nambu-Poisson structures \cite{Nambu:1973qe,Takhtajan:1993vr}, which are a current focus of research (see, e.g., \cite{Park:2008qe,Sato:2010ca,DeBellis:2010sy,Chu:2011yd}) motivated by the pioneering works of \cite{Basu:2004ed,SheikhJabbari:2005mf,Bagger:2006sk,Bagger:2007jr,Gustavsson:2007vu} and others.
Among the early approaches, the one closest to ours is the one
of \cite{Cederwall:1997gg,Bao:2006ef}, which uses $\kappa$-symmetry as guiding principle and features a non-linear self-duality condition. It avoids the use of an auxiliary chiral scalar \cite{Pasti:1997gx} with its covariance problems following a suggestion of \cite{Witten:1996hc}. For these and alternative formulations, e.g., those of
\cite{Howe:1996yn}, based on superspace embedding and $\kappa$-symmetry, we refer to
a recent review \cite{Simon:2011rw}.

\emph{Notation.}---%
Throughout this letter, indices $\al,\be = 0,1,\ldots, p$ label world volume coordinates $\tau = \sigma_0$ and $\vec\sigma = (\sigma_1, \ldots, \sigma_p)$,
$a,b = 1, \ldots, p$ are reserved for the spatial components, $i,j = 0, \ldots, D-1$ denote target space indices, and
capital letters $I, J$ denote ordered \mbox{$p$-tuple} multi-indices  $I = (i_1,\ldots,i_p)$, with $0\leq i_1 < \ldots <  i_p\leq D-1$. A tilde distinguishes fields that carry multi indices. We present our results with Minkowskian signature $(-,+,\ldots,+)$ world volume and target space metric.

\emph{Membrane actions.}---%
A single $p$-brane is described by scalar fields $X^i(\sigma) \equiv X^i(\tau,\vec \sigma)$, $i = 0, \ldots, D-1$, that embed the $(p+1)$-dimensional world volume
$\Sigma$ into the $D$\nobreakdash-\hspace{0pt}dimensional target space manifold $M$. The $p$-brane action in Nambu-Goto form
\begin{equation}\label{NambuGoto}
    S[X] = -T_p \int_\Sigma  \dps \sqrt{-\det (g_{ij} \pa X^i \pb X^j)} \,,
\end{equation}
features the square root of the determinant of the pullback
of the target space metric $g$ to the world volume. (We are particularly interested in the $p=2$, $D=11$ case of the M2 membrane in M-theory/supergravity, but as long as we are discussing the bosonic part of the theory we will keep the world volume and target space dimensions general.)
Introducing an auxiliary world-sheet metric $h_{\al\be}$, we can write a classically equivalent $p$-brane sigma model action
%\footnote{In \cite{Park:2008qe} this action has previously been used to construct an action involving Nambu brackets.}
\begin{multline}\label{membranesigma}
    S[X,h] = -\frac{T_{p}'}{2} \int_\Sigma \dps \sqrt{-\det h} \,\big[g_{ij} h^{\al\be} \pa X^i  \pb X^j \\ - (p-1)\lambda\big] \,,
\end{multline}
where $T_{p}' = \lambda^{\frac{p-1}{2}} T_p$ and  the constant $\lambda > 0$
can be chosen freely. Using the equations of motion of $h_{\al\be}$ one would recover the original action \eqref{NambuGoto}. Instead, we shall
use the reparametrization  invariance of \eqref{membranesigma}  to
locally gauge fix the $h_{a 0}$, $h_{0 b}$ and $h_{00}$ components of the world-sheet metric: $h_{a 0} = h_{0 b} = 0$ for  $a,b = 1,\ldots, p$   and $h_{00} = -\lambda^{p-1}\det (h_{a b})$.
(For world volumes that can be split into a spatial and a temporal part, of the form $\Sigma_p \times \mathbb{R}$, $\Sigma_p \times I$ or $\Sigma_p \times S^1$,
this gauge choice is valid globally.)
In the $p=1$ (string) case, Weyl invariance can be used to gauge fixes all components $h_{\al\be}$, while for  $p>1$, the spatial components $h_{ab}$ are still free. Integrating out  these remaining spatial components of the world-sheet metric (i.e.\ using the equations of motion), we arrive at the gauge-fixed $p$-brane action
\begin{multline}\label{gaugefixed}
    S_\text{gf}[X]  = \frac{T_p}{2} \int \dps \big[ g_{ij} \partial_0 X^i \partial_0 X^j \\ - \det(g_{ij} \partial_a X^i \partial_b X^j)\big] \,.
\end{multline}
It is convenient to rewrite \eqref{gaugefixed} in multi-index notation: Introducing the  antisymmetric $p$-fold tensor product of the target space metric
\begin{equation}\label{gtilde}
    \tilde g_{I\!J} = \sum_{\pi \in \mathfrak{S}_p} \sgn(\pi) g_{i_{\pi(1)} j_1} \cdots  g_{i_{\pi(p)} j_p}
\end{equation}
and the antisymmetric product of partial derivatives
\begin{equation}
\dX^I \equiv \sum_{a_1, \ldots, a_p =1}^p \epsilon^{a_1 \ldots a_p} \partial_{a_1} X^{i_1} \cdots \partial_{a_p} X^{i_p}\,,
\end{equation}
where {\small $i_1 < \ldots < i_p$}, the $p$-brane action becomes
\begin{equation}\label{gfmulti}
S_\text{gf}[X]  = \frac{T_p}{2} \int \dps \big[ g_{ij} \partial_0 X^i \partial_0 X^j  - \tilde g_{I\!J} \dX^I \dX^J\big] \,.
\end{equation}
(It is intriguing to speculate that the ``big'' metric $\tilde g$ in this action could also be independent of $g$.)

Next, we add a background $(p+1)$-form $C_{p+1}$-field
$\frac{1}{(p+1)!}C_{ij_1\ldots j_p} dx^i dx^{j_1} \ldots dx^{j_p}$ with field strength $H = dC$.
Using compact multi-index notation, the corresponding minimal coupling term in the action is
\begin{equation}\label{Cmulti}
    S_C[X]  = - \int \dps \sum_{i,J} C_{iJ} \partial_0 X^i \dX^J \,.
\end{equation}
Eventually, we will also add a $p$-form potential $A$ (describing local fluctuations of the boundary of the membrane) so that all together the gauge invariant combination $C + F$ (locally $F = dA$) enters the theory. We will not consider a dilaton term as it is irrelevant for our discussion.
 For brevity, we will henceforth omit the $p$-brane tension $T_p$ in all formulas.

\emph{Nambu sigma model.}---%
Generalizing the well-known Poisson sigma model \cite{Ikeda:1993fh,Schaller:1994es} to $p>1$, we propose the following Nambu sigma model:
\begin{multline} \label{NambuSigma}
    S[\eta,\tilde \eta, X] = \int \dps \big[ -\frac{1}{2}(G^{-1})^{ij} \eta_i \eta_j + \frac{1}{2} (\tilde G^{-1})^{I\!J} \tilde\eta_I \tilde\eta_J \\
    + \eta_i \partial_0 X^i + \tilde\eta_I \dX^I - \Pi^{i J} \eta_i \tilde\eta_J\big] \,.
\end{multline}
We note the appearance of two types of metrics $G$ and $\tilde G$, as well as %$d$
auxiliary fields $\eta$ and %$d \choose p$ auxiliary fields
$\tilde \eta$.
In the topological case, where $G^{-1}= \tilde G^{-1} = 0$, the consistency of the equations of motion imposes strong conditions on the antisymmetric $(p+1)$-tensor $\Pi$: For $p=1$ consistency requires that~$\Pi$ must be a Poisson tensor. For $p>1$ we find that~$\Pi$  must satisfy the even more stringent conditions of the fundamental identity of a Nambu-Poisson structure \cite{Takhtajan:1993vr}. This differential and algebraic identity has been shown to imply that the multi-vector field $\Pi$ is \emph{decomposable}, i.e.\ it is the wedge product of $p+1$ independent vector fields and it thus defines a $(p+1)$-dimensional submanifold of $M$. Local orthogonal transformations turn $\Pi^{i j_1 \ldots j_p}(x)$ into the Levi-Civita symbol $\epsilon^{i j_1 \ldots j_p}$ (with $i,j_1,\ldots,j_p \in \{0,1,\ldots,p\}$) times a scalar density of weight $-1$ that we shall denote by $|\Pi(x)|^\frac{1}{p+1}$. (For $p=1$ the symbol $|\Pi(x)|$ shall denote the determinant of $\Pi(x)$; for the special case $p+1 = D$ the two possible definitions coincide.) The topological part of the action (\ref{NambuSigma}) agrees with that of an AKSZ construction for topological open p-branes \cite{Bouwknegt:2011vn}.

For non-degenerate $G$ and $\tilde G$ we can use the equations of motion to eliminate the fields $\eta$ and $\tilde\eta$. This yields an action $S[X]$, which  we recognize to be precisely the gauge-fixed action $S_\text{gf}[X] + S_C[X]$ of a $p$-brane in a background $C_{p+1}$-field  (\ref{gfmulti}), (\ref{Cmulti}), with the identifications
\begin{align} 
\begin{split}
g &= (G^{-1} + \Pi \tilde G \Pi^T)^{-1} \\
\tilde g &= (\tilde G^{-1} + \Pi^T  G \Pi)^{-1}\\
C &=  - g \Pi \tilde G = - G \Pi \tilde g  \,.
\end{split}
%\intertext{and, vice versa,} 
%\begin{split}
%G &= g + C \tilde g^{-1} C^T \\
%\tilde G &= \tilde g + C^T g^{-1} C\\
%\Pi &= - g^{-1} C \tilde G^{-1} = - G^{-1} C \tilde g^{-1} \,.
%\end{split}
\end{align}
(Matrix multiplication is understood in all these expression, e.g.\ $C_{jK} = -g_{ji} \Pi^{iJ} \tilde G_{JK}$ with summation over repeated indices.)
We would like to ensure that $\Pi$ is a bona-fide Nambu-Poisson tensor, but for arbitrary background fields $g$, $\tilde g$, and $C$ our present framework is too restrictive for that purpose. We can introduce more freedom into the description by simply splitting $C$ (and $\Pi$) into parts that participate in the transformations above and parts that remain unchanged.
Alternatively, but essentially equivalently, we can write the Nambu sigma model action in block matrix form and augment $G^{-1}$ and $\tilde G^{-1}$,
\begin{equation*}
\begin{pmatrix} -G^{-1} & 0 \\ 0 & \widetilde G^{-1} \end{pmatrix}  \leadsto \begin{pmatrix} -G\quad & \Phi \\ \Phi^T & \widetilde G \end{pmatrix}^{-1} \,,
\end{equation*}
introducing a new $(p+1)$-form field~$\Phi$ that plays a similar role as the corresponding field of \cite{Seiberg:1999vs}. The general open-closed membrane relations that follow from the equivalence of the augmented Nambu sigma model and the $p$-brane action are:
\begin{equation}\label{openclosed}
\begin{split}
g + C \tilde g^{-1} C^T &= G + \Phi \tilde G^{-1} \Phi^T\\
\tilde g + C^T g^{-1} C &= \tilde G + \Phi^T G^{-1} \Phi\\
g^{-1} C  &= G^{-1} \Phi - \Pi (\tilde G + \Phi^T G^{-1} \Phi)\\
C \tilde g^{-1} &= \Phi \tilde G^{-1} - (G + \Phi \tilde G^{-1} \Phi^T)\Pi
\end{split}
\end{equation}
For the special case $p=1$, these relations reduce to the familiar and much simpler open-closed string relations \cite{Seiberg:1999vs}
%(with $\tilde g = g$, $\tilde G = G$, $B:= C$, $\theta:= \Pi$):
\begin{equation} \label{p=1}
\frac{1}{g + C} = \frac{1}{G + \Phi} + \Pi \qquad \text{(for $p=1$)} \vspace{1ex}
\end{equation}

\emph{Nambu gauge theory and Nambu-Poisson map.}---%
We will now add fluctuations in the form of a local world-volume $p$-form gauge potential $A$ with field strength $F = dA$, so that the gauge invariant combination $C + F$ enters the theory. As we shall see, this leads to a new type of gauge theory, which is a higher algebraic analog of noncommutative gauge theory with generalized gauge transformations involving $p-1$-form gauge parameters $\hat \lambda$ and Nambu brackets.

Let us consider the effect of $F$ in the Nambu sigma model.
We have a gauge action of $F$ on $\Pi$ %defined by $F$
\begin{equation} \label{PiF}
\Pi \mapsto \Pi^F = (I-\Pi F^T)^{-1} \Pi = (1 - \langle \Pi,F\rangle)^{-1} \Pi \,,
\end{equation}
where $\langle \Pi,F\rangle \equiv \tr\Pi F^T$.
The second equality in (\ref{PiF}) holds only for $p > 1$ and is due to the decomposability of the Nambu-Poisson tensor $\Pi$. 
(See \cite{Jurco:2000fb,Jurco:2000fs,Jurco:2001my,Jurco:2001kp} for the more complicated case $p=1$.)
$\Pi^F$ is again a Nambu-Poisson tensor. 
%and $\Pi^F$ and $\Pi$ are gauge equivalent.
%For $p=1$ this requires $dF =0$,  For $p > 1$, the gauge equivalence holds in general, but 
Furthermore, for exact $F = dA$, the tensors $\Pi$ and $\Pi^F$ are related by a  Nambu-Poisson map $\rho_{[A]}$ (change of coordinates). To see this, we introduce a tensor $\Pi_t \equiv \Pi^{t F}$ ($t \in [0,1]$) that interpolates between $\Pi$ and $\Pi^F$ and a $t$-dependent vector field $\Pi_t(A) := (\Pi_t)^{iJ} A_J \partial_i$. A straightforward calculation reveals that $\frac{d}{dt} \Pi_t= {\mathcal L}_{\Pi_t(A)} ( \Pi_t) = \langle \Pi_t,F\rangle \Pi_t$. Solving this differential equation, we obtain the geometric series (\ref{PiF}) for $\Pi^F$.
The flow of $\Pi_t(A)$, evaluated at $t=1$, is therefore the Nambu-Poisson map $\rho_{[A]}$ that we are looking for and our construction gives it in closed form. This map is the $p$-brane analog of the (semiclassical) Seiberg-Witten map and it was first obtained in \cite{Chen:2010br}.
Although the Nambu-Poisson tensor $\Pi_t$ does not depend on the
choice of the gauge $p$-potential $A$, the Nambu-Poisson map $\rho_{[A]}$ does: An infinitesimal gauge transformation $\delta A = d \lambda$, induces a change in the flow, which is generated by a vector field $X_{[\lambda, A]} = \Pi^{iJ} (d\hat\lambda_{[\lambda, A]})_J\partial_i$, with a
$(p-1)$-form gauge transformation parameter
\begin{equation}
\hat\lambda_{[\lambda, A]}= \sum_k \frac{(-\mathcal{L}_ {\Pi_t (A)} +
\partial_t)^k(\lambda)}{(k+1)!}\Big| _{t=0}  \,.
\end{equation}
This is the $p$-brane analog of the exact Seiberg-Witten map for the gauge transformation parameter.

Using the Nambu-Poisson map, we can introduce \emph{covariant coordinates} $\hat x^i = \rho_{[A]}(x^i)$. The Jacobian of the transformation $x^i \mapsto \hat x^i$ can be read off \eqref{PiF}, using the decomposability of $\Pi$ for $p>1$.
%\begin{equation}
%\left|\frac{\partial x}{\partial \hat x}\right| = (1 - \langle \Pi,F\rangle) \cdot \left(\frac{|\Pi(x)|}{|\Pi(\hat x)|}\right)^\frac{1}{p+1} \,.
%\end{equation}
The degenerate matrix $F \Pi^T$ acts non-trivially only on a $(p+1)$-dimensional subspace
(via multiplication by~$\langle \Pi,F\rangle$). Altogether,
\begin{equation}\label{jacobian}
\det(1-F\Pi^T) = (1 - \langle \Pi,F\rangle)^{p+1}
= \frac{|\Pi(\hat x)|}{|\Pi(x)|}\cdot\left|\frac{\partial x}{\partial \hat x}\right|^{p+1}  .
\end{equation}

\emph{Effective action.}---%
%The action could be called a $p$-DBI action or a Nambu-Dirac-(Born-Infeld) action or something like that.
We are interested in a system of multiple open M2 branes ending on an M5 brane with a description by an
effective action that is exact to all orders in the coupling constant (for slowly varying fields). Since we focus on the bosonic part of this action, we shall not actually fix $p=2$ and $p'=5$.
Our goal is thus the construction of an effective action for a $p'$-brane with open $p$-branes ending on it while being submerged in a  $C_{p+1}$-background.
The construction is based on two guiding principles: Firstly, this effective action should have dual descriptions similar to the commutative and non-commutative ones of the D-brane and open string case and secondly, it should feature expressions that also appear in the $p$-brane action (\ref{Cmulti}), (\ref{gfmulti}).
%In view of the discussion of the previous paragraphs, the dual descriptions will involve a change of coordinates according to the Nambu-Poisson map.

The open-closed membrane relations (\ref{openclosed}) imply
\begin{multline} \label{miracle}
\det[g + (C+F) \tilde g^{-1} (C+F)^T] =   \\ \det{}^2[1-F\Pi^T]\cdot \det[G+(\Phi + F') \tilde G^{-1} (\Phi + F')^T] \,,
\end{multline}
where $F' = (I - F \Pi^T)^{-1} F$. This miraculous identity holds for all $p$.
Its derivation is quite non-trivial and is best done using block-matrix techniques. It is tempting to take the square root of the identity to construct the action, but recalling \eqref{jacobian}, we notice the appearance of the $2(p+1)$-th power of the Jacobian of the Nambu-Poisson map in \eqref{miracle}. In order to allow dual descriptions and the corresponding change of coordinates, the $2(p+1)$-th root of \eqref{miracle} should enter the effective action that we look for. The Lagrangian density  must be an integral density, we therefore need to complement the part that we have found with an appropriate power of the determinant of the target space metric. These considerations fix the action essentially uniquely and we postulate \vspace{-1ex}
\begin{multline} \label{pDBI}
S_\text{$p$-DBI} =  -\int d^{p'+1} x \, \frac{1}{g_m} \sqrt{-\det g} \\ \cdot \det{}^{\frac{1}{2(p+1)}}\big[1 + (C+F) \tilde g^{-1} (C+F)^T g^{-1}\big] \,,
\end{multline}
where $g_m$ is a ``closed membrane'' coupling constant. The integration is over the larger $p'$-brane and the fields $g$, $\tilde g$, and $C$ in this expression are the pull-backs of the corresponding background target space fields to this $p'$-brane. As desired, the action (\ref{pDBI}) is exactly equal to its ``noncommutative'' dual
\vspace{-1ex}
\begin{multline}\label{pNCDBI}
S_\text{$p$-NCDBI} =  -\int d^{p'+1} x \, \frac{1}{\widehat G_m} \,\frac{\widehat{|\Pi|}^\frac{1}{p+1}}{|\Pi|^\frac{1}{p+1}}  \sqrt{-\det \widehat G} \\ \cdot \det{}^{\frac{1}{2(p+1)}}\big[1 + (\widehat \Phi + \widehat F') \widehat{\tilde G}{}^{-1} (\widehat \Phi + \widehat F')^T \widehat G{}^{-1} \big] \,,
\end{multline}
where $\,\widehat{\;}\,$ denotes objects evaluated at covariant coordinates and $\widehat F'$ is the Nambu (NC) field strength. Vanishing $F$ fixes the open membrane coupling constant
%$G_m$ is related to $g_m$ via
%\begin{equation}
%\frac{G_m}{g_m} = \left(\frac{\det G}{\det g}\right)^{\frac{p}{2(p+1)}} \,.
%\end{equation}
\begin{equation}
G_m = g_m \left(\det G/\det g\right)^{\frac{p}{2(p+1)}} \,.
\end{equation}
The factor involving the quotient of $\widehat{|\Pi|}$ and $|\Pi|$ vanishes for constant $|\Pi|$, but it is essential for the gauge invariance of (\ref{pNCDBI}) in all other cases. As in the string case, we can impose the (background independent) gauge $\Phi = -C$, but we need to restrict the discussion to a maximally noncommutative subspace, where $C$ is non-degenerate. We shall assume that this subspace is $p+1$-dimensional. $\Pi^{iJ}$ and $C_{iJ}$ are square matrices in this subspace and the open-closed membrane relations (\ref{openclosed}) imply $\Pi = -(C^T)^{-1}$, $G = C \tilde g^{-1} C^T$, and $\tilde G = C^T g^{-1} C$. The relevant part of the action (\ref{pDBI}) becomes
%\begin{multline}\label{M}
%S_\text{M} =  \int d^{p+1} x \, \frac{1}{|\Pi|^\frac{1}{p+1}}\,\frac{1}{\widehat g_m} \,  \det{}^{\tt x} (\widehat g) \\
%\cdot \det{}^{\tt y}\big[\widehat\Pi'^T \widehat g \,\widehat\Pi'+ \widehat{\tilde g}{}^{-1} \big] \,.
%\end{multline}
\begin{equation}\label{M}
S_\text{M} =  -\int d^{p+1} x \, \frac{1}{|\Pi|^\frac{1}{p+1}}\,\frac{1}{\widehat g_m} \,  \det{}^{\frac{1}{2(p+1)}}\big[1 + \widehat\Pi'^T \widehat g \,\widehat\Pi'\widehat{\tilde g}\, \big] \,.
\end{equation}
Expanding to lowest order and ignoring a constant we find the infinite-dimensional version of a matrix model
\begin{multline}\label{MM}
-\int d^{p+1} x \, \frac{1}{|\Pi|^\frac{1}{p+1}}\,\frac{1}{2(p+1)\widehat g_m}  \\
\cdot \hat g_{i_0 j_0} \cdots \hat g_{i_p j_p}
\{\widehat X^{j_0},\ldots, \widehat X^{j_p}\} \{\widehat X^{i_0},\ldots, \widehat X^{i_p}\}  \,.
\end{multline}
The physics and mathematics of open $p$-branes ($p>1$) is quite different from that of open strings (\mbox{$p=1$}); nevertheless,
we have been able to choose conventions such that 
all displayed equations except for expressions that involve $\langle\Pi,F\rangle$
%in (\ref{p=1}), (\ref{PiF}), and (\ref{jacobian})
hold for all $1\leq p\leq p'\leq D-1$.

\emph{Relation to previous work and brief discussion.}---%
The DBI part of the M5-brane action in equation (2.9) of \cite{Cederwall:1997gg}
is (modulo conventions)
\begin{equation}
S'=-\int d^6x\, \sqrt{-\det g} \sqrt{1 + \varphi(k)}\, ,
\end{equation}
where $k=(dA+C)_i^{kl}(dA +C)_{jkl}$ and
$\varphi(k) = \frac{1}{3} { \rm tr}k -  \frac{1}{6} { \rm tr}k^2  +
 \frac{1}{36} ({ \rm tr}\,k)^2$.
(See also \cite{Bergshoeff:1996ev}, for an early proposal with a similar index structure.)
The form of the polynomial $\varphi$ has been determined by lengthy computation based on $\kappa$-symmetry and the requirement of
non-linear self-duality. To our surprise, we found that this action $S'$ can be interpreted as a
low-energy (second order in $k$) approximation of our $p$-DBI action (\ref{pDBI}). Indeed,for $p=2$ and $p'=5$ we have $d^{p'+1}x = d^6 x$, $\frac{1}{2(p+1)} = \frac{1}{6}$ and
\[
det^\frac{1}{6}(1 + k) = \sqrt{1 + \frac{1}{3} { \rm tr}k -  \frac{1}{6} { \rm tr}k^2  +
 \frac{1}{36} ({ \rm tr}\,k)^2 + \ldots}\,.
\]
The fact that two very different approaches (one based on $\kappa$-symmetry, the other on commutative/non-commutative duality) give rise to the same action in the low energy limit is very encouraging and seems to indicate that our proposal can indeed be extended to a full supersymmetric action.

%\emph{Discussion.}---%

\emph{Acknowledgments.}---%
It is a pleasure to thank Tsuguhiko Asakawa, Peter Bouwknegt, Chong-Sun Chu, Petr Ho\v rava, Noriaki Ikeda, Matsuo Sato, and Satoshi Watamura for helpful discussions.
%and correspondence.
P.S.\ thanks the particle theory and cosmology group of Tohoku University for hospitality and financial support. The research of B.J.\ was supported by grant GA{\v C}R P201/12/G028.

%\section{Section}
%\subsection{Subsection\label{xx}}
%\subsubsection{Subsubsection}

% If in two-column mode, this environment will change to single-column
% format so that long equations can be displayed. Use
%\begin{widetext}
%\end{widetext}

%\begin{acknowledgments}
%We thank ... for discussion.
%\end{acknowledgments}

%show all possible references:
%\nocite{*}

%\bibliography{nambusigma}


\end{document}